\documentclass[%
 reprint,
 longbibliography,
 amsmath, amssymb,
 aps,
prb,
superscriptaddress
]{revtex4-2}

\usepackage[version=4]{mhchem}
\usepackage{graphicx}
\usepackage{dcolumn}
\usepackage{bm}
\usepackage{physics}
\usepackage{romannum}
\usepackage{hyperref}
\usepackage{color}

\begin{document}


\title{Negligible Kitaev interaction in the van der Waals antiferromagnet NiBr$_2$ revealed by neutron spectroscopy}

\author{Hao Xu}
\affiliation{National Laboratory of Solid State Microstructures and School of Physics, Nanjing University, Nanjing 210093, China} %

\author{Shufan Cheng}
\affiliation{National Laboratory of Solid State Microstructures and School of Physics, Nanjing University, Nanjing 210093, China} %

\author{Yanyan Shangguan}
\affiliation{Department of Applied Physics, Nanjing University of Science and Technology, Nanjing, China} %

\author{Song Bao}
\affiliation{National Laboratory of Solid State Microstructures and School of Physics, Nanjing University, Nanjing 210093, China} %
\affiliation{Collaborative Innovation Center of Advanced Microstructures and Jiangsu Physical Science Research Center, Nanjing University, Nanjing 210093, China}

\author{Junbo Liao}
\affiliation{National Laboratory of Solid State Microstructures and School of Physics, Nanjing University, Nanjing 210093, China} %

\author{Bo Zhang}
\affiliation{National Laboratory of Solid State Microstructures and School of Physics, Nanjing University, Nanjing 210093, China} %

\author{Zihang Song}
\affiliation{National Laboratory of Solid State Microstructures and School of Physics, Nanjing University, Nanjing 210093, China} %

\author{Shuai Dong}
\affiliation{National Laboratory of Solid State Microstructures and School of Physics, Nanjing University, Nanjing 210093, China} %

\author{Maofeng Wu}
\affiliation{National Laboratory of Solid State Microstructures and School of Physics, Nanjing University, Nanjing 210093, China} %

\author{Shinichi Itoh}
\affiliation{Institute of Materials Structure Science, High Energy Accelerator Research Organization,
	Tsukuba, Ibaraki 305-0801, Japan}

\author{Daichi Ueta}
\affiliation{Institute of Materials Structure Science, High Energy Accelerator Research Organization,
	Tsukuba, Ibaraki 305-0801, Japan}

\author{Jinsheng Wen}
\email{jwen@nju.edu.cn}
\affiliation{National Laboratory of Solid State Microstructures and School of Physics, Nanjing University, Nanjing 210093, China} %
\affiliation{Collaborative Innovation Center of Advanced Microstructures and Jiangsu Physical Science Research Center, Nanjing University, Nanjing 210093, China}
\affiliation{Jiangsu Key Laboratory of Quantum Information Science and Technology, Nanjing University, Suzhou, 215163, China}

\date{\today}
\begin{abstract}
	Using time-of-flight neutron spectroscopy, we have mapped out the magnetic excitations in the van der Waals antiferromagnet \ce{NiBr2} over a broad range of energies and momenta. The observed well-defined magnon dispersions are quantitatively reproduced by linear spin-wave theory based on a minimal bilinear Hamiltonian that includes Heisenberg exchange interactions up to third-nearest neighbors, interlayer couplings, and a single-ion anisotropy term, with no need to invoke a Kitaev interaction. This finding is consistent with the fact that bromine is lighter than iodine, leading to weaker ligand-mediated spin-orbit coupling in \ce{NiBr2} than in \ce{NiI2}, and consequently a reduced Kitaev interaction. Our results establish \ce{NiBr2} as a Heisenberg-dominated van der Waals antiferromagnet and support the view that the ligand spin-orbit coupling governs the strength of the bond-dependent interactions within the \ce{NiX2} (X = halogen) family.
\end{abstract}

\maketitle


\section{introduction}

Two-dimensional van der Waals (vdW) magnetic materials have attracted immense research interest in condensed matter physics in recent years. Because they are easily cleavable and structurally highly tunable, these materials offer an excellent highly controllable platform for developing next-generation spintronic~\cite{gongTwodimensionalMagneticCrystals2019,gibertiniMagnetic2DMaterials2019,burchMagnetismTwodimensionalVan2018} and magnonic~\cite{barman2021MagnonicsRoadmap2021,wuSpinExcitationsDynamics2025,chumakMagnonSpintronics2015} devices. Furthermore, the inherent low dimensionality and resulting enhancement of quantum fluctuations, combined with diverse lattice geometries and magnetic anisotropy, make them a uniquely fertile ground for stabilizing exotic quantum states and novel spin interactions~\cite{makProbingControllingMagnetic2019,balentsSpinLiquidsFrustrated2010a, winterBreakdownMagnonsStrongly2017}.
Recently, the realization of bond-dependent anisotropic interactions in vdW materials has emerged as a particularly captivating pursuit, largely inspired by the seminal Kitaev model~\cite{kitaevAnyonsExactlySolved2006}.
Formulated for spin-$1/2$ moments on a honeycomb lattice, the Kitaev model demonstrates how bond-directional exchange can lead to strong magnetic frustration and an exactly solvable quantum spin liquid state.
To realize the Kitaev model in real materials, the conventional approach is to exploit the strong spin-orbit coupling (SOC) of magnetic ions in edge-shared octahedral ligand environments, where it generates spatially anisotropic electronic wavefunctions and thereby enables bond-dependent exchange interactions~\cite{jackeliMottInsulatorsStrong2009a}.
This mechanism has been extensively explored in $4d/5d$ systems~\cite{matsudaKitaevQuantumSpin2025,trebstKitaevMaterials2022}, such as $\alpha$-\ce{RuCl3}~\cite{winterBreakdownMagnonsStrongly2017,ranSpinWaveExcitationsEvidencing2017,maksimovRethinkingAlphaRuCl2020a, banerjeeNeutronScatteringProximate2017,doMajoranaFermionsKitaev2017,yokoiHalfintegerQuantizedAnomalous2021,zhouPossibleIntermediateQuantum2023,czajkaOscillationsThermalConductivity2021} and iridates~\cite{jackeliMottInsulatorsStrong2009a, chaloupkaKitaevheisenbergModelHoneycomb2010,hwanchunDirectEvidenceDominant2015}, as well as in several recently identified Co-based materials~\cite{liuPseudospinExchangeInteractions2018,sanoKitaevHeisenbergHamiltonianHighspin2018,maksimovStrongKitaevInteraction2025,kimAntiferromagneticKitaevInteraction2022}.
However, real materials commonly host additional interactions---{\it e.g.}, Heisenberg exchange and off-diagonal terms---in addition to the Kitaev term, and these perturbations generally cause the ground state to deviate from the ideal Kitaev quantum spin liquid~\cite{chaloupkaZigzagMagneticOrder2013,rauGenericSpinModel2014,winterChallengesDesignKitaev2016,halloranGeometricalFrustrationKitaev2023}.

More recently, the search for Kitaev materials has expanded beyond honeycomb-lattice and $S=1/2$ systems; considerable efforts have been devoted to extending these exotic interactions to a broader range of lattice geometries, including triangular~\cite{kimchiKitaevHeisenbergModelsIridates2014,jackeliQuantumOrderDisorder2015,wangComprehensiveStudyGlobal2021,kimBonddependentAnisotropyMagnon2023} and kagome~\cite{kimchiKitaevHeisenbergModelsIridates2014,ghazanfariTopologicalPhaseTransition2017,moritaGroundstatePhaseDiagram2018} lattices, as well as to higher-spin systems~\cite{stavropoulosMicroscopicMechanismHigherSpin2019,guSignaturesKitaevInteractions2024,shangguanOnethirdMagnetizationPlateau2023,kogaGroundstateThermodynamicProperties2018,zhuMagneticFieldInduced2020,leeTensorNetworkWave2020,khaitCharacterizingSpinoneKitaev2021,xuInterplayKitaevInteraction2018,xuPossibleKitaevQuantum2020,pengKitaevPhysicsTwodimensional2024}.
In particular, theoretical studies on $S=1$ systems have proposed an alternative mechanism in which Kitaev interactions arise not from the SOC of the magnetic ions themselves, but from that of the ligand ions~\cite{stavropoulosMicroscopicMechanismHigherSpin2019,liRealisticSpinModel2023}.  Within this framework, the magnitude of the Kitaev interaction scales approximately with the square of the ligand SOC strength. This mechanism substantially broadens the range of candidate Kitaev materials and predicts that several \ce{Ni^{2+}}-based compounds may host appreciable bond-dependent interactions despite the weak SOC of the magnetic ions.
Among the proposed candidates, the vdW transition-metal dihalide family \ce{NiX2}~($\mathrm{X=Cl,\,Br,\,I}$) provides an ideal platform for testing this theory. These compounds crystallize in the centrosymmetric \ce{CdCl2}-type structure (space group $R\bar{3}m$, No.~166), in which each Ni ion is surrounded by an edge-sharing ligand octahedron~\cite{kuindersmaMagneticStructuralInvestigations1981}. In such an octahedral environment, the \ce{Ni^{2+}} ($3d^8$) ion possesses a quenched orbital angular momentum and a robust $S=1$ ground state~\cite{abragamElectronParamagneticResonance1970}. As a consequence, any Kitaev interaction originating from the SOC of the magnetic ions is expected to be negligible. Therefore, if a sizable Kitaev coupling does exist in these compounds, it must arise predominantly from the ligand SOC, making the \ce{NiX2} family particularly suitable for investigating the ligand-mediated mechanism~\cite{stavropoulosMicroscopicMechanismHigherSpin2019,liRealisticSpinModel2023,kimKitaevInteractionProximate2026,xuSizableLigandmediatedBonddependent2026}.

However, establishing the presence of Kitaev interactions in real materials is a challenging task, since the Kitaev term rarely acts in isolation and is often subdominant to other exchange interactions. Its presence may therefore become apparent only through a subset of unconventional magnetic phenomena,  including anomalous magnon broadening and decay arising from hybridization between the one-magnon branch and the two-magnon continuum, together with a redistribution or suppression of high-energy spectral weight~\cite{kimBonddependentAnisotropyMagnon2023,xieQuantumSpinDynamics2024,xuSizableLigandmediatedBonddependent2026}; magnetization plateaus at fractional values of the saturation magnetization~\cite{shangguanOnethirdMagnetizationPlateau2023}; and an anomalous out-of-plane canting of the magnetic structure~\cite{searsFerromagneticKitaevInteraction2020,liRealisticSpinModel2023,xuSizableLigandmediatedBonddependent2026}. A compelling determination of the Kitaev interaction thus requires mutually consistent constraints from the static magnetic structure, field-dependent properties, and spin dynamics. In particular, spectroscopic measurements with broad energy and momentum coverage are essential for examining not only the magnon dispersion, but also linewidths, and spectral-weight distributions. For NiBr$_2$, previous studies have offered different descriptions of the magnetic interactions, with some invoking a Kitaev term~\cite{jiangKitaevInteractionStabilized2026} and others favouring a predominantly Heisenberg model~\cite{adamNeutronDiffractionStudy1980,regnaultInelasticNeutronScattering1982,dayInelasticNeutronScattering1984}. Establishing whether a Kitaev term is needed to account for the magnetism of \ce{NiBr2} therefore calls for comprehensive time-of-flight inelastic neutron scattering (TOF-INS) measurements.

In this work, we provide the missing constraint for the magnetic interactions by mapping the magnetic excitations of \ce{NiBr2} over a broad range of energy and momentum transfers using TOF INS. The entire measured spectrum is quantitatively reproduced within linear spin-wave theory (LSWT) by a minimal Hamiltonian containing Heisenberg exchanges and SIA, without invoking a Kitaev term. Moreover, we observe neither additional gap features nor anomalous momentum-dependent magnon broadening or spectral-weight suppression of the type expected for an appreciable Kitaev interaction in this noncollinear magnet. By combining these dynamical results with independent constraints from the spin-rotation plane and propagation direction, we show that any Kitaev interaction in bulk \ce{NiBr2} is subdominant and too weak to produce a resolvable effect within our experimental sensitivity. The marked contrast with \ce{NiI2} is consistent with the expected weakening of ligand-mediated bond-dependent exchange from iodine to bromine and provides an experimental benchmark for the ligand-SOC scenario in the \ce{NiX2} family.

\begin{figure}[htbp]
	\includegraphics[width=8.6cm]{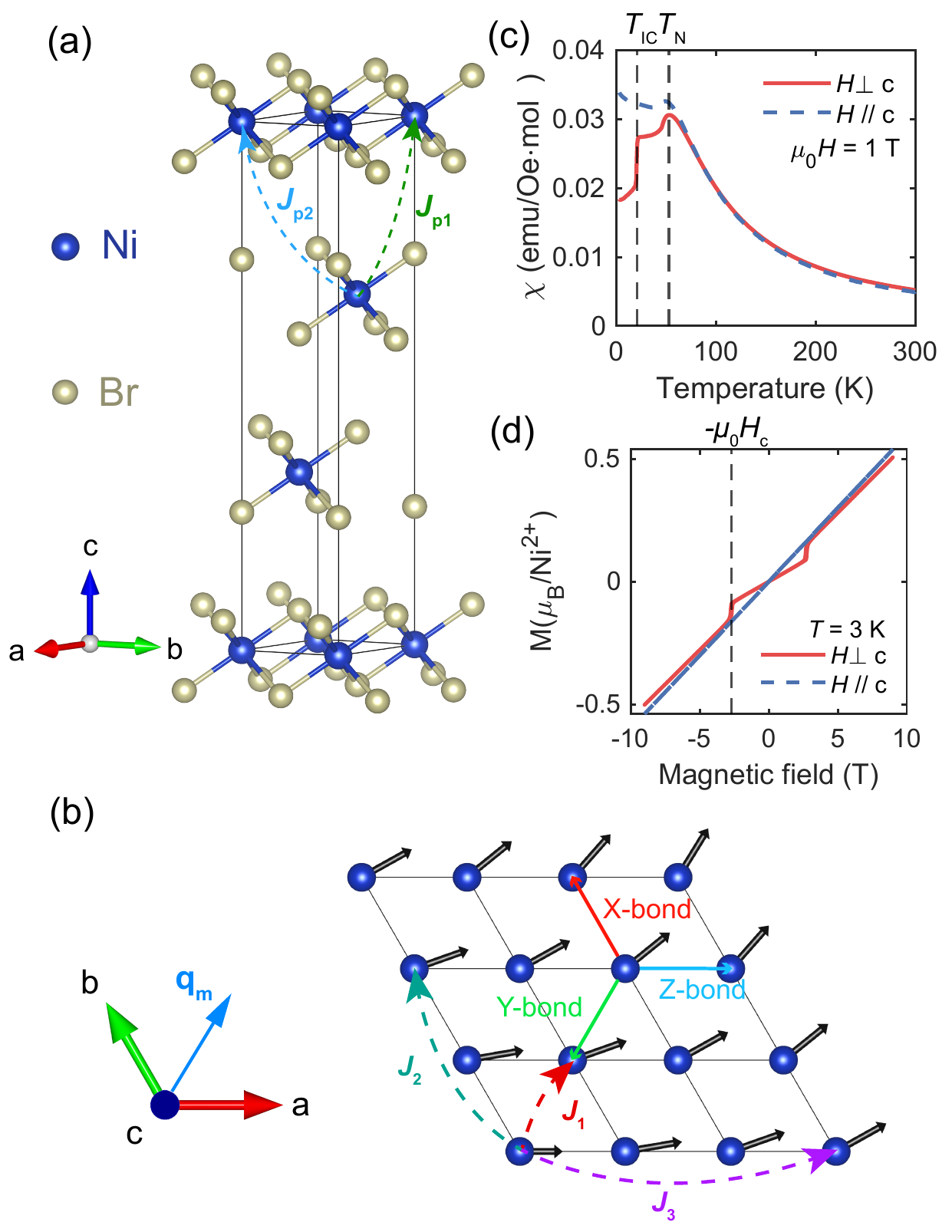}
	\caption{\label{fig: 1} Crystal and magnetic structures, and magnetization of single-crystalline \ce{NiBr2}.
		(a) Schematic of the conventional unit cell, with blue (brown) spheres representing Ni (Br) atoms.
		(b) Magnetic structure of a single layer in the incommensurate phase in the $a$-$b$ plane, with propagation vector $\vb{q}_{\rm m} = (0.027, 0.027, 1.5)$. Black arrows denote spin moments, which lie within the $a$-$b$ plane. Dashed lines in (a) and (b) indicate exchange interaction paths; Kitaev bonds are highlighted by colored arrows.
		(c) Magnetic susceptibility $\chi(T)$ measured under $\mu_0 H = 1$~T applied parallel (dashed) and perpendicular (solid) to the $c$ axis. Vertical lines mark $T_{\rm N} \approx 53$~K and $T_{\rm IC} = 21$~K.
		(d) Field dependence of the magnetization $M$ up to 9~T for $\mu_0 H \parallel$ (dashed) and $\perp$ (solid) the $c$ axis. The vertical line marks the spin-flop transition field $\mu_0H_{\rm c} = 2.7$~T.
	}
\end{figure}

\section{experiment details}

High-quality single crystals of \ce{NiBr2} were successfully grown by the Bridgman method. A total of 2.6~g of \ce{NiBr2} powder (Adamas, 99.99\%) was sealed in a quartz tube, heated to 970~°C, held at this temperature for 110~hours, and then cooled to 850~°C at a rate of 0.5~°C/h. The resulting crystals were brown in color and exhibited shiny cleavage surfaces.

Magnetization measurements were performed using a vibrating-sample magnetometer in a Physical Property Measurement System (PPMS-9T, Quantum Design). INS measurements were carried out on the TOF spectrometer HRC~\cite{itohNeutronBrillouinScattering2013} at the Materials and Life Science Experimental Facility, J-PARC. A 2.4-g single crystal was mounted on an aluminum plate using Cytop grease, fully encapsulated with the same adhesive to prevent hydrolysis, and aligned with the $(HK0)$ plane as the scattering plane.  Data were collected at temperatures of 3, 30, and 300~K, with a primary incident energy $E_{\rm i} = 60$~meV and a chopper frequency of 200~Hz. In the multi-$E_{\rm i}$ mode, this configuration provided two incident energies of 60.9 and 12.42~meV, with corresponding energy resolutions of 3.46 and 0.36~meV at the elastic line, respectively. The sample was rotated through 120° in steps of 0.5°. Data analysis was performed using the HORACE software package~\cite{ewingsHoraceSoftwareAnalysis2016}, and the INS intensity was corrected by dividing by the Bose factor. Theoretical spectra were calculated within the LSWT framework using the SpinWaveTheorySpiral routine implemented in the Sunny.jl package~\cite{dahlbomSunnyjlJuliaPackage2025}.

\section{results}

\begin{figure*}[t]
	\includegraphics[width=17.5cm]{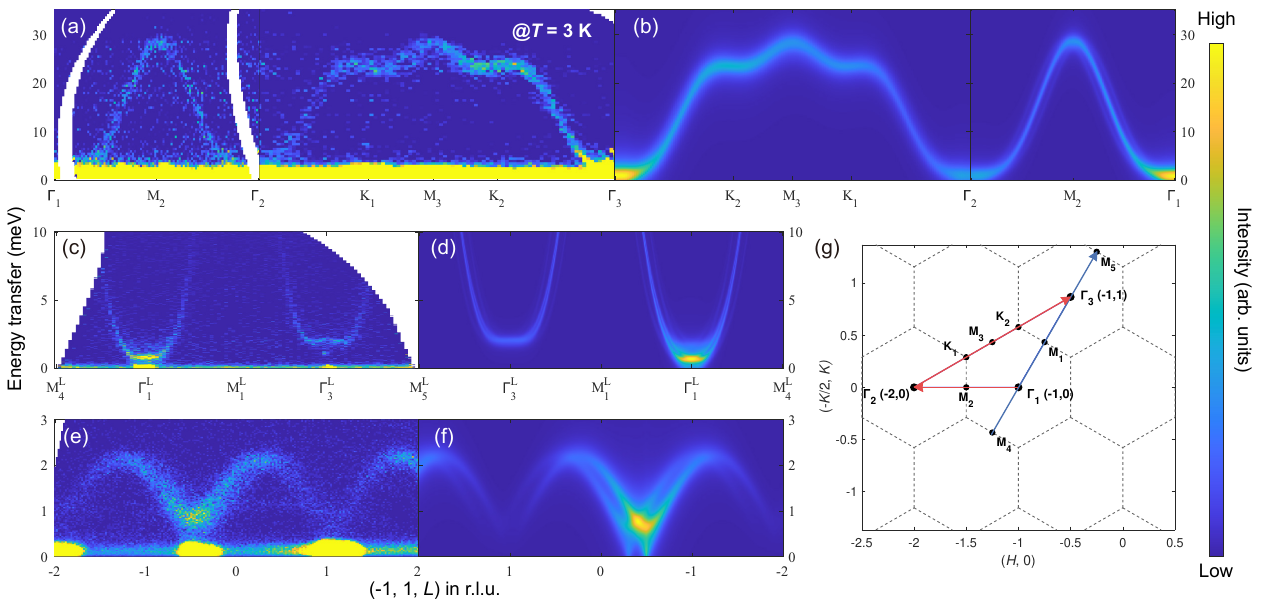}
	\caption{\label{fig: 2} INS data and LSWT calculations at $T = 3$~K in the IC phase of \ce{NiBr2}.
		(a) INS intensity along the red path shown in (g). Owing to the weak dispersion along $L$ [see panel (e)], the data are integrated over $L \in [-1,1]$. The in-plane data are integrated over $\Delta K \in [-0.1,0.1]$ along $(-0.5K, K)$ for the path $\Gamma_1 \to \Gamma_2$, and over $\Delta K \in [-0.1,0.1]$ along $(-K, K)$ for $\Gamma_2 \to \Gamma_3$.
		(b) Corresponding LSWT dispersions along $\Gamma_3 \to \Gamma_2 \to \Gamma_1$.
		(c) INS intensity along the blue path shown in (g), integrated over $\Delta K \in [-0.05,0.05]$ along $(-0.5K, K)$ and $\Delta L \in [-0.1,0.1]$ along the $L$ direction. The superscript $L$ in the high-symmetry labels denotes $\bar{L} = 0.5$.
		(d) Corresponding LSWT dispersions along $M_5^L \to M_4^L$.
		(e) INS intensity along $(-1,1,L)$, integrated over $\Delta H \in [-0.07,0.07]$ along $(H,0)$ and $\Delta K \in [-0.07,0.07]$ along $(-0.5K, K)$.
		(f) Corresponding LSWT dispersions along $L$.
		(g) Two-dimensional reciprocal lattice and high-symmetry points in the $(H, K)$ plane. Red and blue lines indicate the paths of INS data.
		For comparison with the experimental data, the calculated dispersions are averaged over two crystallographic domains and integrated over the same region of the Brillouin zone as in the experiment. See SM~\cite{SeeSupplementalMaterial} for more details.
	}
\end{figure*}

\subsection{Magnetic Susceptibility}

Figure~\ref{fig: 1}(c) shows the temperature dependence of the magnetic susceptibility of a single-crystal sample under a magnetic field $\mu_0 H = 1$~T applied parallel and perpendicular to the $c$ axis. Both the in-plane and out-of-plane susceptibilities exhibit a broad peak at $T_{\rm N} \approx 53$~K, indicating an antiferromagnetic (AFM) transition. In this intermediate AFM phase, spins within each layer form ferromagnetic sheets, while adjacent layers are coupled antiferromagnetically, with a propagation vector $\vb{q}_{\rm m} = (0,\,0,\,1.5)$~\cite{adamNeutronDiffractionStudy1980}.
Upon cooling below $T_{\rm IC} = 21$~K, the in-plane susceptibility (red solid line) shows a sharp drop, indicative of a first-order phase transition. The magnetic structure evolves into an incommensurate (IC) helical state. As illustrated in Fig.~\ref{fig: 1}(b), spins within each layer lie in the basal plane with an in-plane propagation vector $(0.027,0.027,0)$. Spins in adjacent layers remain antiferromagnetically coupled, yielding a total propagation vector $\vb{q}_{\rm m} = (0.027,\,0.027,\,1.5)$~\cite{adamNeutronDiffractionStudy1980}. Meanwhile, the lattice undergoes a slight distortion~\cite{adamNeutronDiffractionStudy1980}; however, due to its small magnitude, all reciprocal lattice vectors are indexed using the high-temperature hexagonal unit cell throughout this work.

Analysis of the high-temperature inverse susceptibility in the range of 200-300~K shows that $\chi^{-1}$ follows the Curie-Weiss law, $\chi^{-1} = (T - \Theta_{\rm p})/C$ [see Fig.~S1 in the Supplemental Material (SM)~\cite{SeeSupplementalMaterial}]. The Curie-Weiss temperatures are determined to be $\Theta_{\rm p} = 48.64$ and 56.32~K for the in-plane and out-of-plane directions, respectively, indicating that ferromagnetic (FM) interactions within the plane are dominant in this system. The effective magnetic moments, extracted from the slopes, are $\mu_{\rm eff} = 3.24$ and 3.09~$\mu_B$ for the in-plane and out-of-plane data, respectively.
In \ce{NiBr2}, the magnetic \ce{Ni^{2+}} ions are coordinated by an octahedral ligand field, which leads to a quenching of the orbital angular momentum and stabilizes a pure spin $S = 1$ ground state~\cite{abragamElectronParamagneticResonance1970}. Based on the extracted effective moments, the corresponding $g$ factors are estimated to be 2.29 and 2.18, both close to the ideal spin-only value $g_{\rm s} = 2$. This is consistent with previous microwave resonance experiments, where the in-plane $g$ factors are reported to be 2.16 in the high-temperature paramagnetic phase~\cite{wellsEPRSingleCrystals1966}, 1.94 in the intermediate phase~\cite{katsumataAntiferromagneticResonanceNiBr21969}, and 2.06 in the IC phase~\cite{adamMagneticResonanceExperiments1980}. All of these results consistently support an $S = 1$ spin-triplet ground state in \ce{NiBr2}.
As shown in Fig.~\ref{fig: 1}(d), we further investigate the field dependence of the magnetization in the IC phase. A clear spin-flop transition~\cite{raiMagneticPropertiesFesubstituted2022} is observed at an in-plane critical field of $\mu_0 H_{\rm c} = 2.7$~T, which provides additional evidence for an easy-plane spin anisotropy.

\subsection{INS Spectra}

Figures~\ref{fig: 2}(a), (c), and (e) display INS spectra at $T = 3$~K in the IC phase along different high-symmetry paths in the Brillouin zone [Fig.~\ref{fig: 2}(g)]. A well-defined magnon branch is observed, featuring an in-plane band top of 28.5~meV at the $M$ points, a shoulder at 23.5~meV near the $K$ points, and an out-of-plane band top of 2.1~meV at $(-1,1,0.5+n)$ with $n \in \mathbb{Z}$.
Given the elastic-line energy resolution of 0.36~meV for $E_i=12.42$~meV, only data above 0.5~meV are considered. Within this range, the INS intensity exhibits a local maximum at 0.88~meV at $\Gamma_1^{L}$, which likely corresponds to the band bottom of the dispersion.

Based on these spectra, we construct a minimal bilinear spin Hamiltonian for \ce{NiBr2}. The easy-plane anisotropy, inferred from the magnetic structure determined by neutron diffraction and supported by the magnetization data in Fig.~\ref{fig: 1}(c) and (d), is incorporated as a symmetry-allowed SIA term $A_{zz}(S_i^z)^2$ with $A_{zz} > 0$.
Isotropic Heisenberg exchange interactions are included up to third-nearest neighbors within the plane. As suggested by previous studies~\cite{liRealisticSpinModel2023,liEffectsKitaevInteraction2024b}, the incommensurate magnetic order arises from the competition between FM nearest-neighbor (NN) exchange $J_1$ and AFM third-NN exchange $J_3$.
Interlayer interactions are included up to second-nearest neighbors. This is motivated by previous studies on the sister compound \ce{NiI2}~\cite{xuSizableLigandmediatedBonddependent2026,liRealisticSpinModel2023,kimKitaevInteractionProximate2026}, where the second-nearest-neighbor coupling $J_{\rm p2}$ plays a dominant role and may also be relevant in the present system.
The Kitaev interaction is not included in the present model, primarily because the magnetic excitation spectrum does not exhibit the anomalous features observed in \ce{NiI2}, such as the pronounced broadening and suppression of magnon spectral weight associated with strong magnon-magnon interactions. Instead, the excitations in \ce{NiBr2} remain well-defined magnon modes throughout the measured energy and momentum range, suggesting that the Kitaev interaction in this system is likely very weak, even if it is present. A more detailed discussion of the Kitaev term will be provided in Sec.~\ref{sec: discussion}.
For now, we consider the following Hamiltonian:

\begin{equation}
	\begin{aligned}
		H = & \sum_{\expval{i, j}} J_1 \vb{S}_i \cdot \vb{S}_j +\sum_{\expval{\expval{i, j}}} J_2 \vb{S}_i \cdot \vb{S}_j
		+ \sum_{\expval{\expval{\expval{i, j}}}} J_3 \vb{S}_i \cdot \vb{S}_j                                              \\
		+   & \sum_{\expval{i, j}^\perp} J_{\rm p1} \vb{S}_i \cdot \vb{S}_j
		+ \sum_{\expval{\expval{i, j}}^\perp} J_{\rm p2} \vb{S}_i \cdot \vb{S}_j
		+ A_{zz} \sum_i ({S}_i^z)^2
	\end{aligned}
	\label{Eq: J1J2J3 model}
\end{equation}
where $\expval{i, j}$, $\expval{\expval{i, j}}$ and $\expval{\expval{\expval{i, j}}}$ denote pairs of first, second and third NN within layers, while $\expval{{i, j}}^{\perp }$ and $\expval{\expval{i, j}}^{\perp }$ refer to the first and second-NN interlayer couplings. The exchange pathways are shown in Fig.~\ref{fig: 1}(a) and (b).

The magnetic structure of the model is examined using the classical Luttinger-Tisza method, which requires $A_{zz} > 0$ and imposes a constraint on the exchange parameters~\cite{SeeSupplementalMaterial}.
Then, by calculating the spin-wave excitation using the Sunny.jl package~\cite{dahlbomSunnyjlJuliaPackage2025} and fitting the INS data [see SM~\cite{SeeSupplementalMaterial} for details], we obtain a set of optimal parameters, as listed in Table~\ref{table: J1J2J3}. Figures~\ref{fig: 2}(b), (d), and (f) show the corresponding LSWT spectra for comparison with the experimental intensity. With these optimal parameters, the calculated dispersions well reproduce the observed features, including the overall magnon dispersion, the energies at high-symmetry points, and the spectral intensities.

We further vary each parameter around its optimal value and monitor the corresponding changes in the spectra (see Fig.~S6 in the SM~\cite{SeeSupplementalMaterial}).
This analysis reveals the distinct roles of the individual interactions. The nearest-neighbor exchange $J_1$ primarily controls the overall bandwidth of the in-plane magnon dispersion. While $J_2$ and $J_3$ also modify the bandwidth, they respectively leave the magnon energies at the high-symmetry $K$ and $M$ points nearly unchanged. The single-ion anisotropy $A_{zz}$ mainly determines the gap at the bottom of the dispersion. Both interlayer couplings, $J_{\rm p1}$ and $J_{\rm p2}$, primarily control the bandwidth of the out-of-plane dispersion and produce qualitatively similar effects. In addition, they also influence the in-plane dispersions because the vectors connecting the nearest- and second-nearest-neighbor interlayer interactions possess finite in-plane components.

\begin{table}[htbp]
	\caption{Summary of the optimal parameters of Eq.~\ref{Eq: J1J2J3 model} and their corresponding uncertainties.}
	\begin{ruledtabular}
		\begin{tabular}{ccc}
			Parameter    & Optimal Value (meV) & Uncertainty (meV) \\
			\colrule
			$J_1$        & -3.3667             & 0.1334            \\
			$J_2$        & -0.0826             & 0.1319            \\
			$J_3$        & 0.9999              & 0.0236            \\
			$A_{zz}$     & 0.1001              & 0.1240            \\
			$J_{\rm p1}$ & 0.1612              & 0.2029            \\
			$J_{\rm p2}$ & 0.2076              & 0.1786            \\
		\end{tabular}
		\label{table: J1J2J3}
	\end{ruledtabular}
\end{table}

\begin{figure}[htbp]
	\includegraphics[width=8.6cm]{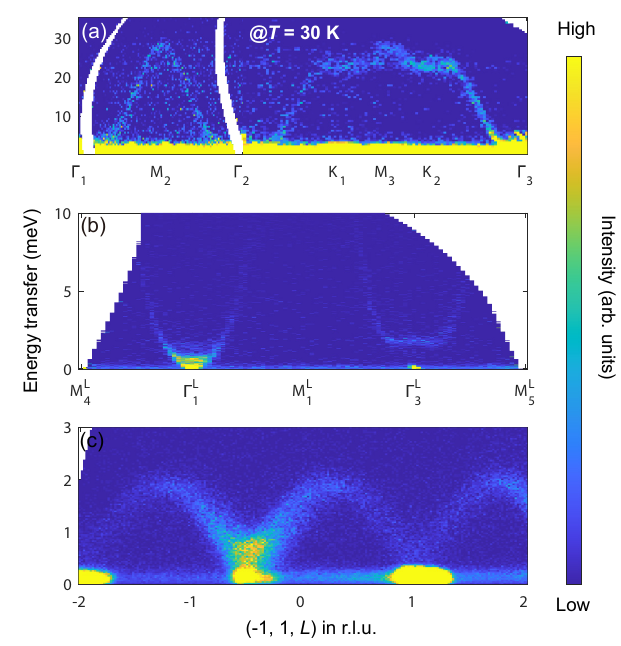}
	\caption{\label{fig: 30K}
		INS data at $T = 30$~K in the intermediate AFM phase of \ce{NiBr2}. All the k-paths and integration ranges are the same as those described in Fig.~\ref{fig: 2}.
	}
\end{figure}

In addition to the IC phase, we have also examined the INS data in the intermediate AFM phase at 30~K. As shown in Fig.~\ref{fig: 30K}, the magnon dispersions in the AFM phase are essentially identical to those in the IC phase, indicating that the underlying exchange interactions do not change significantly across the transition. Meanwhile, since the in-plane incommensurate propagation vector $(0.027,\,0.027,\,0)$ in the IC phase is very small, the shift of magnetic Bragg peaks has only a negligible effect on the excitation spectra.

\section{Discussion}\label{sec: discussion}

The analysis above demonstrates that the magnetic excitations in \ce{NiBr2} are well described by a minimal Hamiltonian comprising Heisenberg exchange interactions and SIA. We now consider additional interactions that may be present, focusing on the Kitaev and biquadratic exchange terms and assessing their possible influence on the magnetic ground state and excitation spectrum.

\begin{figure}[htbp]
	\includegraphics[width=8.6cm]{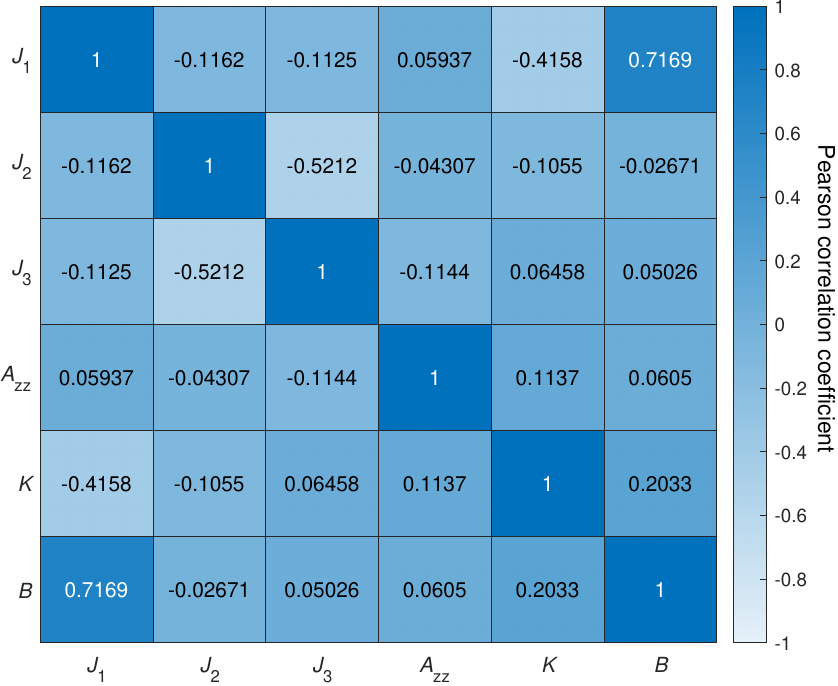}
	\caption{\label{fig: correlation}
		Correlation matrix for the six fitted parameters in the extended model including $K$ and $B$. The analysis includes parameter sets with $\chi^2$ values within 10\% of the minimum; see the SM~\cite{SeeSupplementalMaterial} for details.
	}
\end{figure}

We first examine how these interactions modify the calculated excitation spectrum. As shown in Fig.~S7~\cite{SeeSupplementalMaterial}, both the Kitaev interaction $K$ and the biquadratic interaction $B$ affect the spectrum in a manner qualitatively similar to that of $J_1$. To quantify this behavior, we perform additional fits to the INS data using an extended Hamiltonian that explicitly includes both $K$ and $B$. To limit the number of free parameters and avoid overparameterization, the interlayer couplings $J_{\rm p1}$ and $J_{\rm p2}$ are fixed at the values listed in Table~\ref{table: J1J2J3}, whereas all other parameters are allowed to vary. The resulting optimal parameters are summarized in Table~\ref{table: KB}. As shown by the parameter-convergence curves in Fig.~S8~\cite{SeeSupplementalMaterial}, the cost functions for $K$ and $B$ are relatively flat near their minima, resulting in large uncertainties. We further perform a correlation analysis using parameter sets with $\chi^2$ values within 10\% of the minimum. The resulting correlation matrix, shown in Fig.~\ref{fig: correlation}, reveals strong linear correlations of both $K$ and $B$ with $J_1$, consistent with their qualitatively similar effects on the excitation spectrum. In the following, we combine these fitting results with constraints from the excitation spectrum and magnetic structure to assess the magnitude and physical relevance of each interaction, while also clarifying the origins of the observed parameter correlations.

We begin by considering the Kitaev interaction. Although \ce{NiBr2} has been proposed as a candidate for ligand-spin-orbit-coupling-mediated Kitaev interactions, our TOF INS results, together with previous experimental and theoretical studies, consistently indicate that the Kitaev interaction in bulk \ce{NiBr2} is weak. First, no gap other than that associated with SIA is resolved within the experimental energy resolution. The Kitaev term can be decomposed into an isotropic part, $(K/3)\mathrm{I}_{3\times3}$, and a traceless anisotropic component, $K_{\rm anis}$. The isotropic part effectively renormalizes $J_1$ and thereby modifies the overall magnon bandwidth, whereas the anisotropic component can lift band degeneracies and open gaps at high-symmetry points. For the easy-plane magnetic ground state of \ce{NiBr2}, the observed gap at $(-1,1,-0.5)$ is predominantly generated by SIA, and no additional gap attributable to Kitaev anisotropy is resolved. Moreover, any small gap induced by a weak Kitaev interaction is likely obscured by the experimental and computational broadening. Consequently, among the parameter sets with $\chi^2$ values within 10\% of the minimum, the primary effect of $K$ is to renormalize $J_1$, accounting for the strong correlation between these two parameters.

Second, \ce{NiBr2} exhibits well-defined, long-lived magnon modes throughout the Brillouin zone, characteristic of conventional spin-wave behavior. This observation indicates that the anharmonic effects associated with Kitaev interactions~\cite{kimBonddependentAnisotropyMagnon2023,xieQuantumSpinDynamics2024} are weak. The spectrum contrasts sharply with that of \ce{NiI2}, where pronounced magnon damping over a broad region of momentum space and a substantial suppression of high-energy spectral weight have been attributed to coupling between one-magnon excitations and the multimagnon continuum. Nevertheless, the $S=1$ character of \ce{NiI2} suppresses quantum fluctuations, and its excitation spectrum remains largely describable within LSWT despite its sizable Kitaev interaction. In the $S=1/2$ triangular-lattice van der Waals magnet \ce{CoI2}~\cite{kimBonddependentAnisotropyMagnon2023}, by contrast, Kitaev-induced anharmonicity reconstructs the excitation spectrum well beyond the LSWT one-magnon picture, with the spectral boundary largely determined by the two-magnon continuum. These three compounds therefore illustrate a systematic reduction in Kitaev-induced anharmonic effects from \ce{CoI2} to \ce{NiI2} and finally to \ce{NiBr2}.

Third, the magnetic structure determined by neutron diffraction provides an independent constraint on the Kitaev interaction. The spin-rotation plane in \ce{NiBr2} coincides with the $a$-$b$ plane, indicating that the magnetic anisotropy is dominated by easy-plane SIA. In systems with quenched orbital angular momentum, magnetic anisotropy determines the orientation of the spin-rotation plane. The weak Kitaev interaction in \ce{NiBr2} allows SIA to confine the spins to the $a$-$b$ plane. By contrast, in \ce{NiI2}, the sizable Kitaev interaction cants the spins away from the basal plane, producing a spin-rotation plane tilted by approximately $55^\circ$ from the $c$ axis~\cite{liRealisticSpinModel2023,xuSizableLigandmediatedBonddependent2026,congSoftMagnonsVan2024}. The in-plane propagation direction also provides a sensitive measure of the Kitaev interaction. Placing the fitted parameters in the previously established phase diagram of the $S=1$ triangular-lattice Kitaev-Heisenberg model~\cite{liEffectsKitaevInteraction2024b} shows that, in the parameter regime relevant to \ce{NiBr2}, an antiferromagnetic Kitaev interaction can readily rotate the propagation vector from the experimentally observed $[h, h]$ direction to $[h,0]$. Based on the phase boundaries reported in Ref.~\cite{liEffectsKitaevInteraction2024b}, we conservatively estimate an upper bound of approximately 1~meV for the Kitaev interaction in \ce{NiBr2}.

In addition to the Kitaev term, DFT calculations suggest a biquadratic interaction $B(\vb{S}_i\cdot\vb{S}_j)^2$ in the \ce{NiX2} ($\mathrm{X=Cl, Br, I}$) family~\cite{niGiantBiquadraticExchange2021,liRealisticSpinModel2023}. Although higher-order interactions are traditionally assumed to be small, recent studies have shown that they can substantially affect magnetic ground states and excitation spectra~\cite{wysockiConsistentModelMagnetism2011,moreno-cardonerCaseStudyUniaxial2014,milaOriginBiquadraticExchange2000,zhuGiantBiquadraticInteractioninduced2016,kaoKitaevInteractionsVan2026a}. Nevertheless, accurately extracting $B$ from INS spectra within LSWT is intrinsically difficult because, at quadratic order, the biquadratic interaction generates a bilinear contribution that is algebraically equivalent to the conventional Heisenberg term. This can be seen directly from the Holstein--Primakoff expansion. Writing
\begin{equation}
	\vb{S}_i\cdot\vb{S}_j=\mathcal{C}+\hat{\mathcal{H}}_{ij}^{(2)}+\mathcal{O}(a^3),
\end{equation}
where $\mathcal{C}$ is the classical contribution and $\hat{\mathcal{H}}_{ij}^{(2)}$ contains terms quadratic in the bosonic operators, gives
\begin{equation}
	(\vb{S}_i\cdot\vb{S}_j)^2\approx \mathcal{C}^2+2\mathcal{C}\hat{\mathcal{H}}_{ij}^{(2)}+\mathcal{O}(a^3).
\end{equation}
Thus, within LSWT, the leading effect of $B$ is indistinguishable from a renormalization of $J_1$, explaining the particularly strong correlation between these parameters. Determining the biquadratic interaction therefore requires a probe sensitive to physics beyond the one-magnon sector. Resonant inelastic x-ray scattering, for example, can in principle access multipolar excitations characteristic of higher-order spin interactions. In practice, however, the biquadratic interaction is expected to be much weaker than the Heisenberg exchanges, making the required energy resolution experimentally challenging.

\begin{table}[htbp]
	\caption{Optimal parameters and corresponding uncertainties for the extended model including the Kitaev interaction $K$ and biquadratic interaction $B$.}
	\begin{ruledtabular}
		\begin{tabular}{ccc}
			Parameter    & Optimal value (meV) & Uncertainty (meV) \\
			\colrule
			$J_1$        & -3.0515             & 0.3010            \\
			$J_2$        & 0.0474              & 0.1339            \\
			$J_3$        & 0.9148              & 0.0802            \\
			$A_{zz}$     & 0.1384              & 0.0130            \\
			$K$          & 0.3735              & 0.9253            \\
			$B$          & 0.7993              & 1.2964            \\
			$J_{\rm p1}$ & 0.1612              & fixed             \\
			$J_{\rm p2}$ & 0.2076              & fixed             \\
		\end{tabular}
	\end{ruledtabular}
	\label{table: KB}
\end{table}

\section{Conclusion}

In this work, we characterize the temperature and field dependence of the magnetic properties of single-crystalline \ce{NiBr2} using magnetization measurements and investigate the magnetic excitation spectra in both the IC phase and the intermediate AFM phase using TOF INS. By modeling the magnon excitations within LSWT and performing a detailed fit to the INS spectra in the IC phase, we establish a minimal bilinear Hamiltonian comprising Heisenberg exchange interactions and SIA, without the need to invoke a Kitaev interaction. This conclusion is supported by the well-defined magnon excitations throughout the Brillouin zone, the absence of additional gap features beyond those associated with the easy-plane anisotropy, and the experimentally observed spin-rotation plane and in-plane propagation direction. These observations are consistent with the substantially weaker SOC of Br than that of I, which results in a much weaker ligand-mediated Kitaev interaction in \ce{NiBr2} than in \ce{NiI2}.
Taken together, our results provide strong experimental support for the ligand-SOC-mediated origin of Kitaev interactions in the \ce{NiX2} family.

\section*{Acknowledgement}

The work was supported by the National Key Projects for Research and Development of China (Grants No.~2024YFA1409200 and No.~2021YFA1400400), the National Natural Science Foundation of China (Grants No.~12434005, No.~12225407, No.~12622409, and No.~12404173), the Natural Science Foundation of Jiangsu Province (Grants No.~BK20233001, No.~BK20241251, and No.~BK20241250), the Xiaomi Young Scholars Technology Innovation Award, and the Fundamental Research Funds for the Central Universities (Grants No.~KG202501 and No.~14380262). The neutron experiment at the Materials and Life Science Experimental Facility of the J-PARC was performed under a user program (Proposal No.~2025A0040).


%

\end{document}